\documentclass[useAMS,usenatbib]{mn2e}

\usepackage{verbatim}
\usepackage{graphicx}

\title[Modelling stellar coronae from surface magnetograms: the
role of missing magnetic flux]{Modelling stellar coronae from surface magnetograms: the
role of missing magnetic flux}

\author[Johnstone et al.]{C. Johnstone$^1$, M. Jardine$^1$, D. H. Mackay$^2$ \\ $^1$ School of Physics and Astronomy, Universtiy of St Andrews, St Andrews, Scotland KY16 9SS \\ $^2$ School of Mathematics and Statistics, University of St Andrews, North Haugh, St Andrews, Scotland Fife KY16 9SS}

\begin{document}

\maketitle

\begin{abstract}
Recent advances in spectropolarimetry have allowed the reconstruction of stellar coronal magnetic fields. This uses Zeeman-Doppler magnetograms of the surface magnetic field as a lower boundary condition. The ZDI maps, however, suffer from the absence of information about the magnetic field over regions of the surface due to the presence of dark starspots and portions of the surface out of view due to a tilt in the rotation axis. They also suffer from finite resolution which leads to small scale field structures being neglected.

This paper explores the effects of this loss of information on the extrapolated coronal fields. For this we use simulated stellar surface magnetic maps for two hypothetical stars. Using the potential field approximation, the coronal fields and emission measures are calculated. This is repeated for the cases of missing information due to, (i) starspots, (ii) a large area of the stellar surface out of view, (iii) a finite resolution.

The largest effect on the magnetic field structure arises when a significant portion of the stellar surface remains out of view. This changes the nature of the field lines that connect to this obscured hemisphere. Nonetheless, the field structure in the visible hemisphere is reliably reproduced. Thus the calculation of the locations and surface filling factors of accretion funnels is reasonably well reproduced for the observed hemisphere. The decrease with height of the magnetic pressure, which is important in calculating disc truncation radii for accreting stars, is also largely unaffected in the equatorial plane. The fraction of surface flux that is open and therefore able to supply angular momentum loss in a wind, however,  is often overestimated in the presence of missing flux.

The magnitude and rotational modulation of the calculated emission measures is consistently decreased by the loss of magnetic flux in dark starspots. For very inactive stars, this may make it impossible to recover a magnetic cycle in the coronal emission. Finite resolution has little effect on those quantities, such as the emission measure and the average coronal electron density, that can currently be observed.
\end{abstract}

\section{Introduction}

Since magnetic activity was first detected on stars other than the Sun \citep{1947ApJ...105..105B}, a great deal has bean learned about the structure of stellar magnetic fields and their effects on the evolution of stars. Using the Zeeman effect on magnetically sensitive spectral lines, it is possible to measure the mean surface fields (using Zeeman splitting/broadening) and the mean line-of-sight fields (using circular polarization) for unresolved stars. 

Significant progress has been made since it has become possible to identify individual structures on the surfaces of unresolved stars as they rotate. This is done using a technique called Doppler-imaging which allows surface temperature and chemical composition maps of rapidly rotating stars to be reconstructed from rotationally broadened spectral lines \citep{1983ApJ...275..661V}. Doppler-imaging was extended by \citet{1989A&A...225..456S} to a technique called Zeeman-Doppler imaging (ZDI) which uses the Stokes I and circularly polarized Stokes V parameters to give reliable surface magnetic maps (e.g. \citealt{1997MNRAS.291....1D}, \citealt{2006Sci...311..633D}, \citealt{2008MNRAS.390..567M}).

This technique has been used with great success to study the kinds of surface magnetic fields that many stars possess. For example, \citet{2006MNRAS.370..629D} produced ZDI maps for the massive, slowly rotating star $\tau$ Sco. It was found that $\tau$ Sco possesses a strong, complex, mostly poloidal field that is distributed over all latitudes and constant over the period of 1.5 years that the star was observed. These observations are inconsistent with current dynamo models for massive stars. They concluded that $\tau$ Sco probably possesses a fossil-field.

For some studies, such as those involving the magnetic interaction between a young star and its disc, knowledge of the 3-dimensional structure of a star's magnetic field is also required. Making a few simplifying assumptions, and using ZDI maps as a lower boundary condition, the structure of the coronal fields can be extrapolated. The most common technique that has been used is the potential-field source surface (PFSS) model developed by \citet{1969SoPh....9..131A}, described below. 
    
Recently, ZDI maps for four classical T Tauri stars have shown that they possess complex fields that depart significantly from dipolar configurations (\citealt{2007MNRAS.380.1297D}, \citealt{2008MNRAS.386.1234D}, \citealt{2009arXiv0905.0914H}). This is significant as the field is thought to interact with circumstellar discs in ways that can depend on what kind of field the stars have. This interaction can have a large effect on the accretion of material in circumstellar discs onto stellar surfaces. Magnetospheric accretion models for classical T Tauri stars usually assume that the star possesses a dipolar magnetic field (\citealt{1990RvMA....3..234C}, \citealt{1991ApJ...370L..39K}). More complex fields, however, can affect several aspects of accretion models. These effects include the locations of hotspots where accretion columns, funneled along field lines, impact onto the stellar surface (\citealt{2006MNRAS.371..999G}) and angular momentum transfer between discs and their stars (\citealt{2008MNRAS.386.1274L}, \citealt{2009ASPC..406..112R}).

Significant complexity is also typically found in the magnetic fields of main sequence stars that have been observed with this technique. One of the stars that has been most extensively studied in this way is the rapidly rotating, active star, AB Dor (e.g. \citealt{1997MNRAS.291....1D}, \citealt{1999MNRAS.302..437D}, \citealt{1999MNRAS.305L..35J}, \citealt{2003MNRAS.345.1145D}). \citet{1999MNRAS.305L..35J} and \citet{2000ASPC..198..463H} found that this technique used on observed radial field maps can accurately reproduce the observed azimuthal field maps up to a latitude of $60^\circ$. \citet{1999MNRAS.305L..35J} suggested that the disagreement above $60^\circ$ could be due to the presence of dark polar spots suppressing the Zeeman signature at these latitudes. They also noted that due to a tilt of about $60^\circ$ to our line of sight in the star's rotation axis, a large part of the surface is out of view, and thus all studies on this star must be carried out without any information over about an entire third of the surface. 

While the ZDI technique has been applied with great success, allowing for the global structure of magnetic fields to be explored in stars other than the Sun, several fundamental difficulties arise from the finite ability of the ZDI maps in picking up field structures. 

As stars are unresolved, differentiating between magnetic features above and below the equator, when using the Doppler imaging and ZDI techniques, requires that the rotation axis of the stars be tilted to a certain extent. Unfortunately, this means that ZDI maps are unable to reproduce the field over the entire surface of any star. 

Another problem comes from the presence of dark regions on the surface of stars that are generally assumed to be the stellar analogue of sunspots. If this assumption is valid, then they would presumably contain strong magnetic fields with strengths that exceed the rest of the stellar surface. \citet{2002AN....323..165S} considers the validity of this assumption. \citet{2005LRSP....2....8B} argues that the differences between observed starspot filling factors and magnetic filling factors, along with the fact that regions of intermediate brightness have been observed to contain the strongest magnetic fields on ZDI maps, indicates that the fields from starspot umbrae have yet to be detected. If this is the case, it may be significant limitation to ZDI maps and subsequently to the reconstruction of the 3-dimensional fields. 

The surface field maps also suffer from having limited spatial resolution. Since ZDI maps actually measure magnetic flux averaged over each surface element, regions of opposite polarity will cancel each other out and thus not be picked up on the map if they are not resolved. The field strengths derived from Zeeman-Doppler Imaging are always lower than those derived from Zeeman broadening. Any structures not present on ZDI maps will obviously not be accounted for when the field is extrapolated to higher radii.

The aim of this paper is to explore some of the limitations that these effects impose on extrapolating global coronal fields. Starting from the surface fields of two simulated stars, where these problems are not present, the global coronal fields and the emission measures are calculated. The effects discussed above are simulated so that the calculations can be repeated and compared.

\section{Model}
\begin{figure}
\includegraphics[width=84mm]{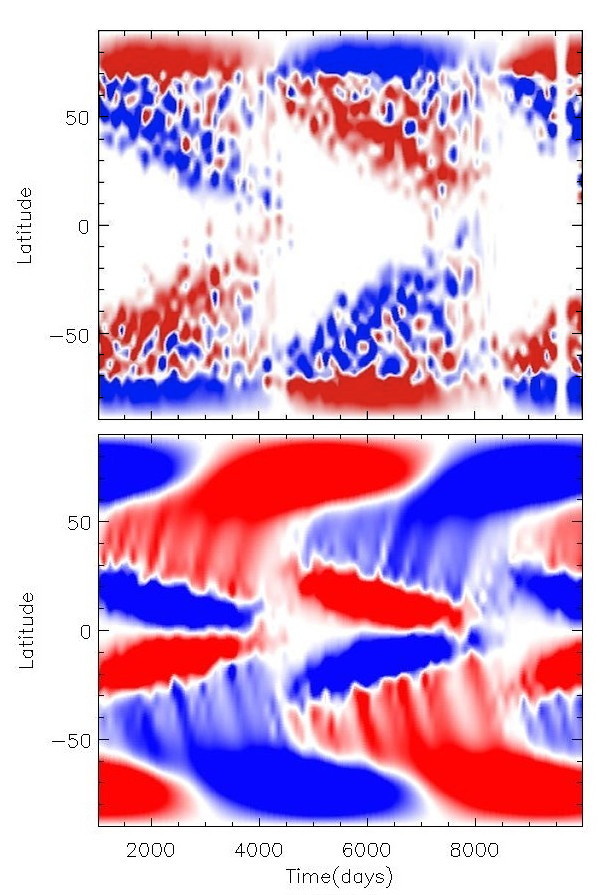}
\caption{Simulated butterfly diagrams for an active star (top) and an inactive star (bottom) showing the signed magnetic flux integrated over all longitudes as a function of latitude and time. Both diagrams show almost two and a half 11-year activity cycles. Red and blue regions represent net fluxes of opposite polarity. \label{butterfly}}
\end{figure}
\begin{figure}
\includegraphics[width=84mm]{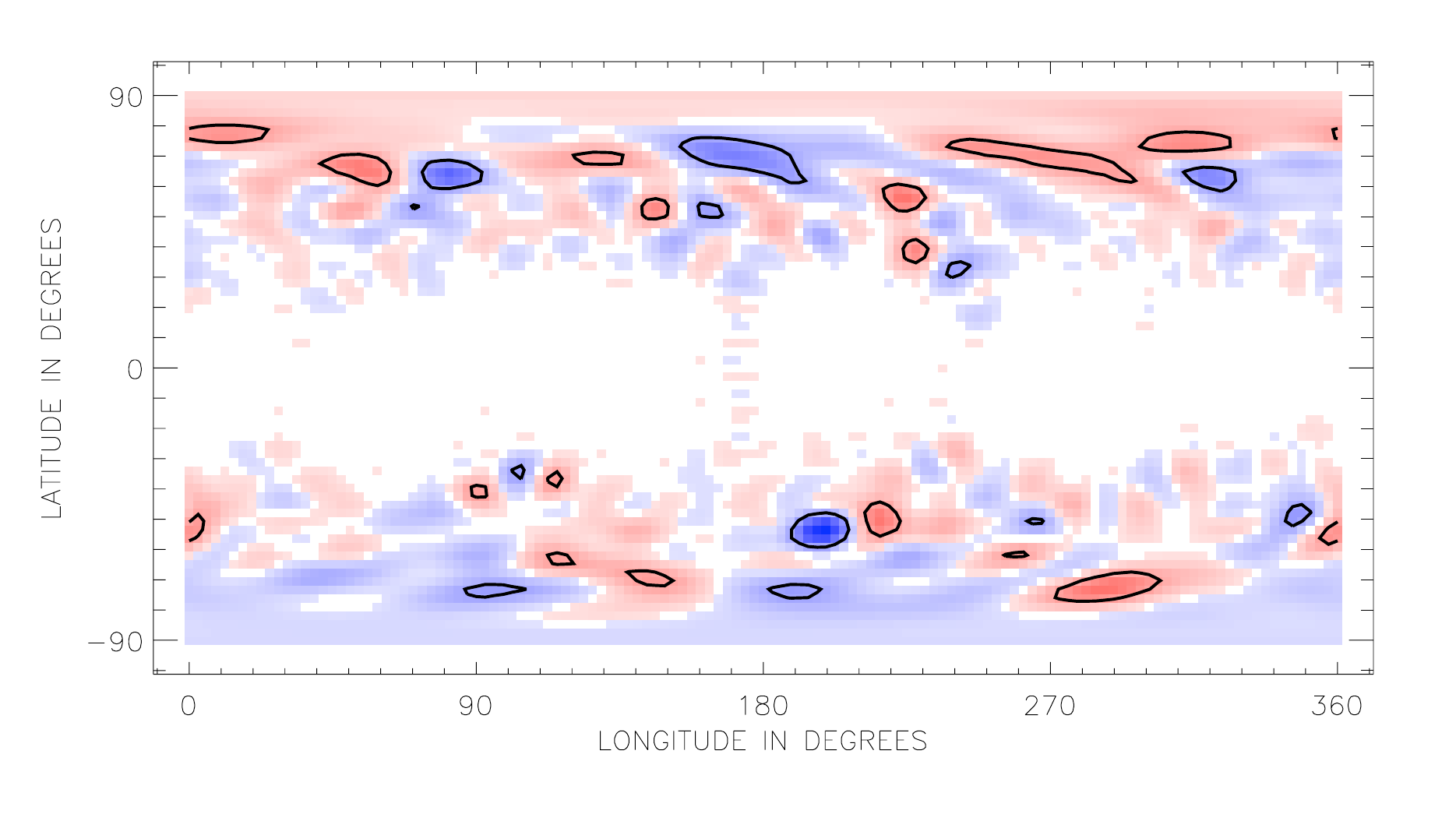}
\includegraphics[width=84mm]{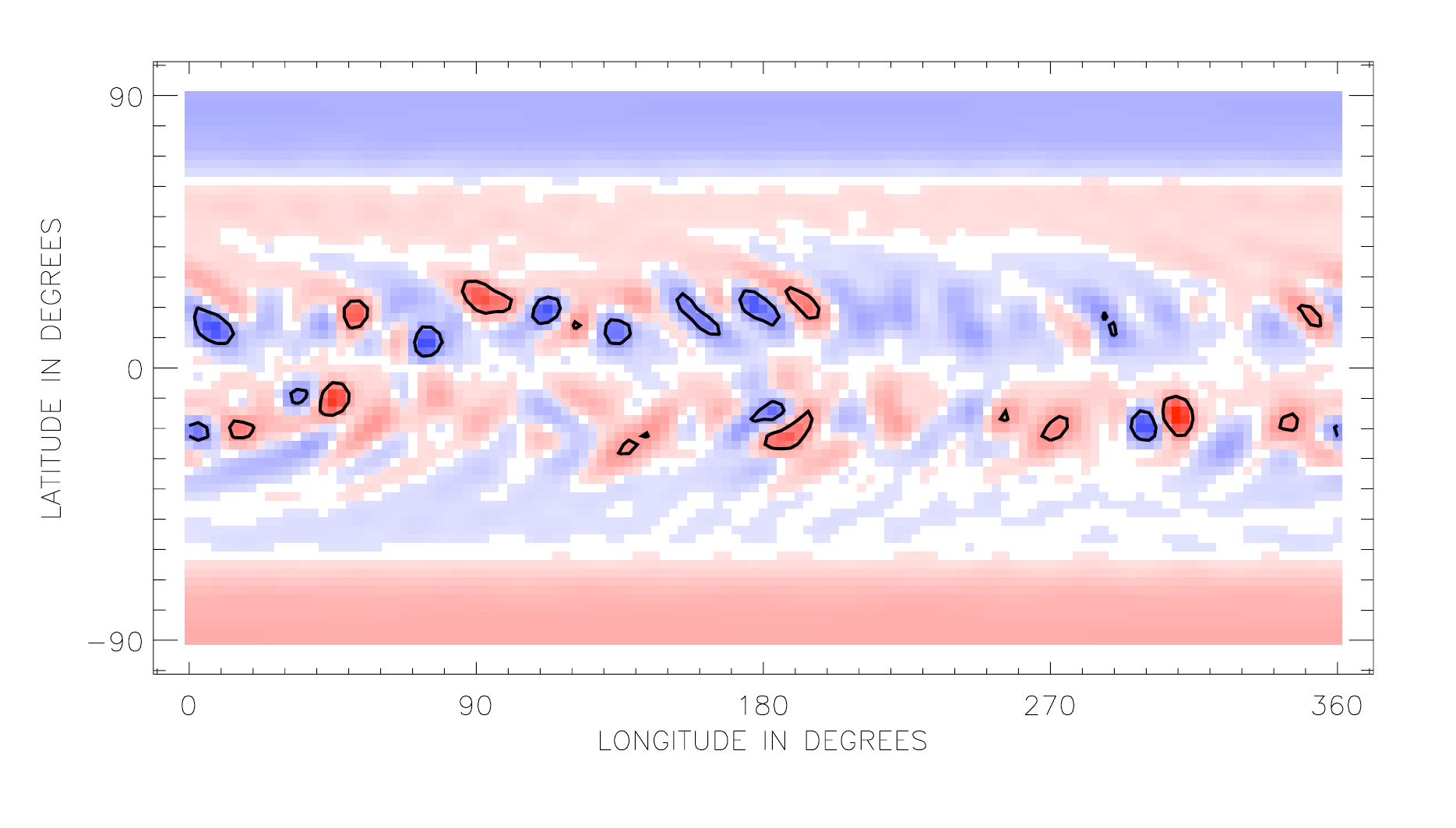}
\caption{Surface radial field maps for an active star (top) and an inactive star (bottom) at a time of a relative maximum in magnetic activity. Black lines give contours at 500G. Red and blue regions represent magnetic fields of opposite polarity. \label{surfacemaps}}
\end{figure}

\subsection{Simulated Stellar Magnetograms}

The radial component of the magnetic field at the base of the stellar corona is simulated using the magnetic flux transport model of \citet{2004MNRAS.354..737M} and \citet{1998ApJ...501..866V}. The model describes the time evolution of the radial component of the magnetic field at the stellar surface through the magnetic induction equation under the effects of differential rotation, meridional flow, supergranular diffusion and the emergence of new magnetic fields in the form of magnetic bipoles.

This paper will use two simulations of stars from \citet{2004MNRAS.354..737M}. These stars differ by the range of latitudes at which flux is emerged and the meridional flow. In both cases, a comparable level of magnetic flux is emerged. The star given the greater meridional flow and range of latitudes for flux emergence (hereinafter the "active star") has flux emerging between latitudes of $10^\circ$ and $70^\circ$ in both hemispheres and a meridional flow of $100 m s^{-1}$. The other star (hereinafter the "inactive star") has flux emerging between latitudes of $10^\circ$ and $40^\circ$ in both hemispheres and a meridional flow of $11 m s^{-1}$. For both stars, the diffusion of magnetic flux is characterized by the diffusion constant, $D$ (see Equation 1 of  \citealt{2004MNRAS.354..737M}), which is given the value of $450 km^2 s^{-1}$. The activity of the active star is analogous to the young active star AB Dor while that of the inactive star is analogous to the Sun.

The magnetic activity of both stars is simulated to give periodic variations over 11-year cycles. Butterfly diagrams showing the latitudinal concentration of magnetic activity over these cycles for both stars are given in Fig. \ref{butterfly}. As can be seen, the concentrations of polar flux visible in Fig. \ref{surfacemaps} are present throughout the cycles of both stars. Both stars show reversals in polar flux between every cycle. Sample radial field surface maps are given in Fig. \ref{surfacemaps}. These correspond to times of maximum magnetic activity for both stars. Black contour lines are drawn at field strengths, $| B_r |$, of 500G.

\subsection{Potential-Field Extrapolation}\label{sec:potentialmodel}

The simulated surface fields are extrapolated to higher radii using the potential-field source surface model described by \citet{1969SoPh....9..131A}. The field is assumed to be current-free ($\nabla \times \bf{B} = 0$) and thus satisfies Laplace's equation ($\nabla^2 \Phi = 0$). The 3D coronal field can then be described as the sum over multiple spherical harmonic components and is given by the following equations where $P_{lm}$ are the associated Legendre polynomials.

\[
B_r = - \sum_{l=1}^{N} \sum_{m=-1}^{l} [ l a_{lm} r^{l-1} - (l+1) b_{lm} r^{-(l+2)}] P_{lm} (\cos \theta) e^{i m \phi}
\] 

\[
B_\theta = - \sum_{l=1}^{N} \sum_{m=-1}^{l} [  a_{lm} r^{l-1} + b_{lm} r^{-(l+2)}] \frac{d}{d\theta} P_{lm} (\cos \theta) e^{i m \phi}
\] 

\[
B_\phi = - \sum_{l=1}^{N} \sum_{m=-1}^{l} [  a_{lm} r^{l-1} + b_{lm} r^{-(l+2)}] P_{lm} (\cos \theta) \frac{im}{\sin \theta} e^{i m \phi}
\] 

As an upper boundary condition, we assume that the field becomes radial at a {\it source surface} which is taken to be at a radius of $3.4 R_*$. With this, the radial component of the field becomes

\begin{equation}\label{eq:field}
B_r (r, \theta, \phi)= \sum_{l=1}^N \sum_{m=0}^{l} c_{lm} f(r,l) P_{lm} (\cos \theta) e^{im\phi}
\end{equation}

where

\[
f(r,l)=\left(\frac{(l+1)(\frac{r}{R_{*}})^{-(l+2)}+l(\frac{r_s}{R_{*}})^{-(2l+1)}(\frac{r}{R_{*}})^{l-1}}{l+1+l(\frac{r_s}{R_{*}})^{-(2l+1)}} \right).
\]

The surface radial field maps are used as a lower boundary condition to derive the values of $c_{lm}$ for all values of $l$ and $m$ under consideration. For this study, values of $l$ up to 31 will be considered. The field is extrapolated using a modified version of the code developed by \citet{1998ApJ...501..866V}. 

Once the global field is determined, the coronal gas pressures can be estimated using the model described in \citet{2002MNRAS.333..339J} and \citet{2006MNRAS.371..999G}. The gas pressure in regions corresponding to open field lines is taken to be zero and thus only regions of the stellar corona confined by closed magnetic loops contribute to the emission measure. The gas pressure at the stellar surface is assumed to be proportional to the magnetic pressure ($p_0 \propto B_0^2$) and to vary along field lines assuming the gas is in hydrostatic equilibrium, given by

\[
p=p_0 {\rm exp} \left[\frac{m}{kT} \int g_s ds \right]
\]
where $g_s$ is the component of effective gravitational acceleration along the field line. Once a temperature is assumed, the density $n_e$ can then be determined. We assume an emissivity proportional to $n_e^2$ to obtain the X-ray emission measure. We consider both its magnitude and rotational modulation throughout the stellar magnetic cycle.

\subsection{The effects of  star spots, the inclination of the stellar rotation axis and finite surface resolution}

Once the coronal magnetic field structure, gas pressures and emission measures have been calculated using the full surface magnetograms, the process is repeated using magnetograms that have been degraded to mimic the effects of a) the masking of the Zeeman signature in dark starspots that leads to a reduced magnetic flux measurement b) the lack of information from a hemisphere that is out of view because of the inclination of the stellar rotation axis and c) a finite surface resolution. 

The presence of star spots is simulated by cutting from the maps all regions with fields, $| B_r|$, stronger than a threshold of 500G. 
Similarly, the obscuration of a portion of the star's surface due to a tilt in its rotation axis is simulated by cutting out an entire hemisphere. This corresponds to a limiting case where the star's rotation axis is parallel with our line of sight and we can only see one hemisphere. If a star actually had this tilt angle, the features on its surface would not be Doppler shifted as the star rotates, and thus stellar magnetograms would be impossible to produce. Even though this does not correspond to a physically realistic situation, we nonetheless consider it as it provides the most severe limitation on the fraction of the star that is visible. 
We note however, that while the very early ZDI maps were produced by reconstructing the radial component, or all three vector components, of the field at each point on the surface independently (\citealt{1991A&A...250..463B}, \citealt{1999MNRAS.302..437D}), the more recent reconstructions (\citealt{2001MNRAS.326.1265D}, \citealt{2006MNRAS.370..629D}, \citealt{2008MNRAS.384...77M}) describe the field in terms of spherical harmonics whose amplitudes are fitted to the line profiles. This allows the reconstruction of magnetic modes that are symmetric or antisymmetric about the equator and therefore some indication of the global symmetry of the field can be obtained by comparing the fit to the data of the modes, even if some portion of one hemisphere is out of view.

The opposite limiting case, where the star's axis of rotation is perpendicular to our line of sight, also makes it impossible for ZDI maps to be produced. In this case, features at the same latitude in opposite hemispheres are indistinguishable in the broadened spectra and so it is impossible to assign magnetic elements to a single hemisphere.

The effect of having low-resolution ZDI maps is simulated by smearing the surface field such that the original maps, with a latitudinal resolution of just over $1^\circ$ and an equal longitudinal resolution at the equator, are replaced by maps with latitudinal resolution of around $11^\circ$ degrees and a longitudinal resolution of around $8^\circ$ at the equator. As ZDI maps measure magnetic flux, each new surface element contains a field which is simply the area weighted mean of the fields in the old elements contained within it.

In all three of these cases, the magnetic field is extrapolated and the results compared to the original case. In analysing the difference between the results, two aspects will be considered. These are the effects on the structure of the magnetic field and the effects on the coronal gas pressures, densities and emission measures. Also, the effects on the open flux and the field lines that can potentially support accretion from a circumstellar disc are explored.

\section{Results}

\subsection{The surface maps}

\begin{table}
\begin{tabular}{|c|ccc|}
\hline 
Star & $\Delta \Phi_{\textrm{spots}}$ (\%) & $\Delta \Phi_{\textrm{hemisphere}}$ (\%) & $\Delta \Phi_{\textrm{smeared}}$ (\%)\\
\hline
Active & 19 & 52 & 17 \\
Inactive & 11 & 51 & 10 \\
\hline
\end{tabular}
\caption{Percentage drop, averaged over one activity cycle, in total magnetic flux, $\Phi$, over the stellar surface for both stars. This drop is due to the suppression of magnetic field in dark starspots and an entire hemisphere out of view and the cancellation of opposite polarity magnetic fields due to the smearing of the stellar surface magnetogram.}
\label{tbl:flux}
\end{table}

\begin{figure*}
\includegraphics[width=170mm]{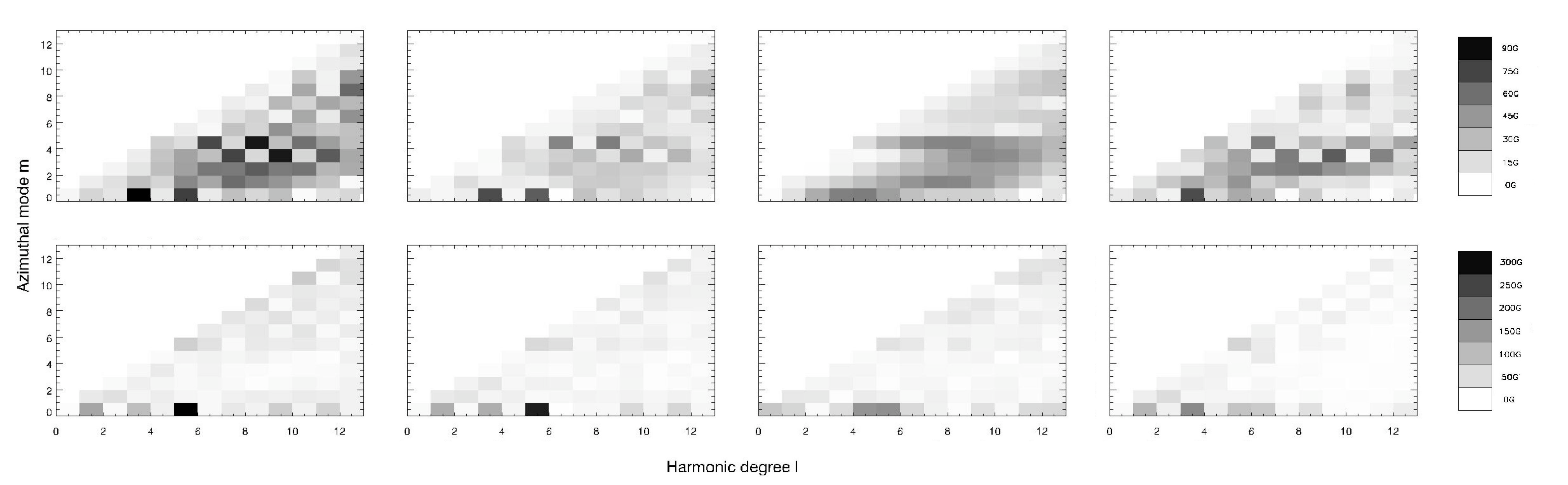}
\caption{Plots showing the strength of each $c_{lm}$ coefficients given in Equation \ref{eq:field}. Results are shown for the full data case (first column), the field cutoff above 500G (second column), the field cutoff in an entire hemisphere (third column) and the field smeared (forth column) for both the active star (top) and inactive star (bottom).\label{structure}}
\end{figure*}

The effect of degrading the surface magnetograms can be seen very clearly by plotting the amplitudes $c_{lm}$ of the spherical harmonic components from equation (\ref{eq:field}) as a function of $l$ and $m$. This is shown in Fig. \ref{structure} for a period of maximum activity during the magnetic cycle. This corresponds to the same time as that shown in the magnetograms in Fig. \ref{surfacemaps}. The first column shows the distribution of power across the modes for both the active star (top) and the inactive star (bottom). The magnetic field of the active star shows power across a much greater range of modes, corresponding to its more complex and widely-distributed field. The second column, illustrating the effect of dark star spots, shows the influence on the field structure of cutting out all regions of field above 500G. This clearly has a greater effect on the field of the more active star since it has a greater coverage of these strong-field regions. While some power is removed from all the modes, it is the high-order modes that are most strongly suppressed, while the dominant low-order modes are largely unaffected. The third column illustrates the effect of removing field from the unobserved hemisphere. This has a significant effect on the field structures of both stars. It allows power to be distributed more evenly over the different modes, giving the effect of blurring the $c_{lm}$ plot given in the second column of Fig. \ref{structure}. The fourth column illustrates the effect of smearing the original magnetogram. This removes some of the higher order components while leaving the lower order components unchanged. As the strength of higher order components decreases with increasing radius faster than lower order components, this affects the field on smaller scales more than on larger scales. 

Table \ref{tbl:flux} shows the percentage drop (over the stellar activity cycle) in the total unsigned surface magnetic flux for both the active and inactive star. The case where an entire hemisphere is cut from the surface magnetograms presents the largest drop in magnetic flux of about $50 \%$ on average. For the active star, the case where the surface magnetogram is smeared gives a large drop in magnetic flux due to cancellation of opposite polarity flux. 

Although these stars show large unsigned magnetic flux over the entire magnetogram, according to Gauss' law for the magnetic field ($\oint_S \mathbf{B} \cdot d\mathbf{S} =0$) the total signed flux must always be zero (in the absence of magnetic monopoles). However, when regions of the magnetograms are suppressed, if more flux of one polarity is lost, this condition will not be satisfied. This is the case when the fields stronger than 500G and when the field in an entire hemisphere are suppressed, although in both cases, the total signed flux is usually smaller than 2\% of the total unsigned flux. In these calculations, the surface magnetograms are represented as the sum over many spherical-harmonic components. The total signed flux for each individual component is zero and thus for the spherical-harmonic expansion of the surface magnetograms is also zero. Thus, the extrapolated magnetic fields, in most cases, cannot perfectly represent the original surface magnetograms when regions of the magnetograms are suppressed. Instead of being zero, the total unsigned flux in the hemisphere that is suppressed after the extrapolation, is usually around 5 - 10 \% of its original (unsuppressed) value.

\subsection{The magnetic field structure, accretion and open flux} \label{sect:fieldstructure}

\begin{figure*}
\includegraphics[width=55mm]{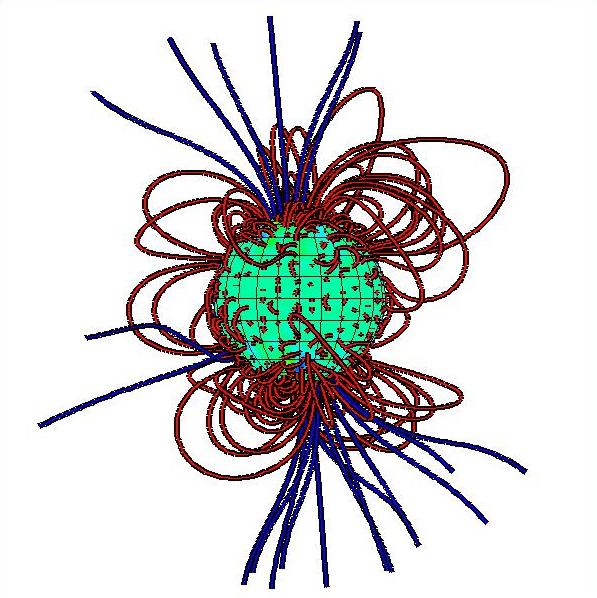}
\hspace{20mm}
\includegraphics[width=55mm]{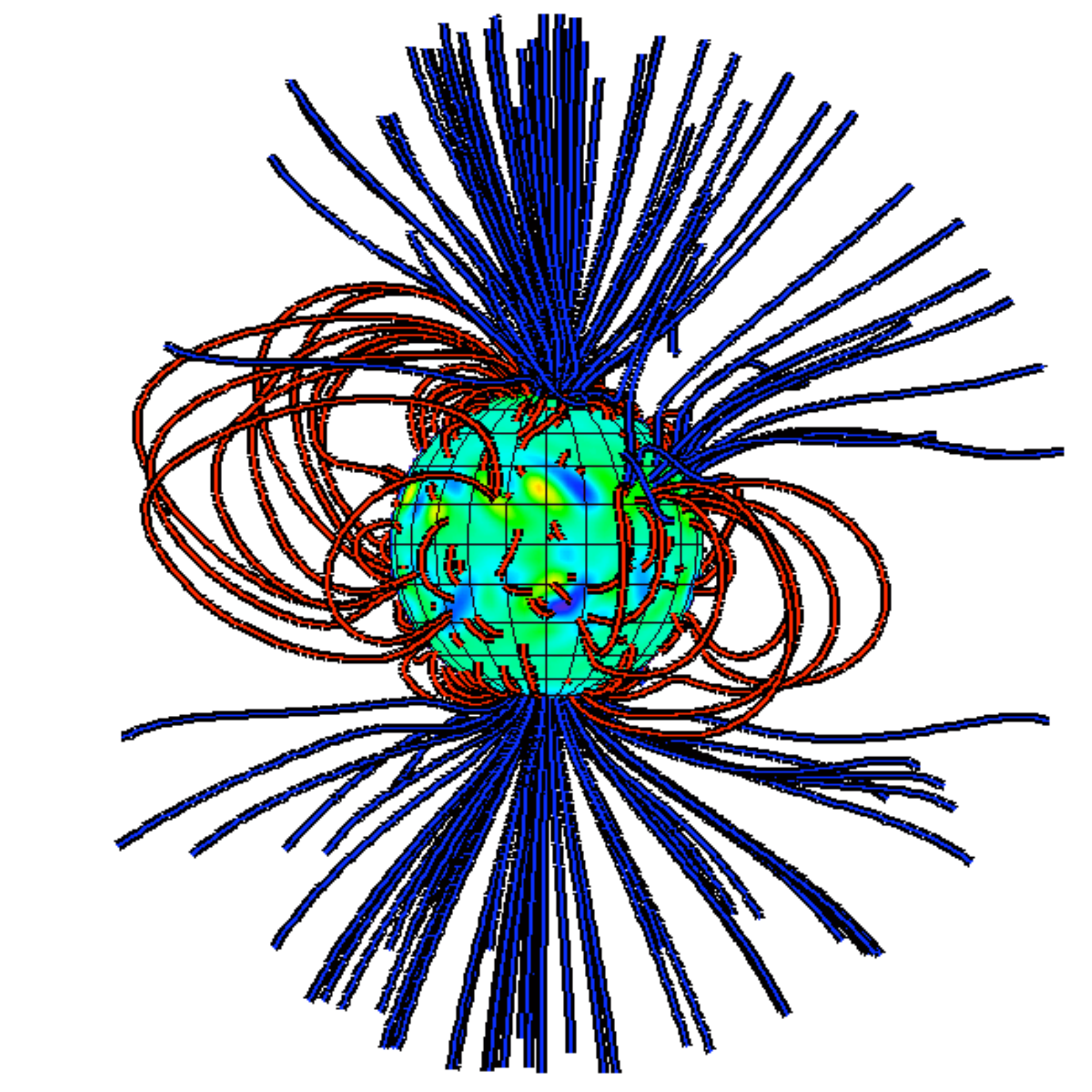}
\includegraphics[width=55mm]{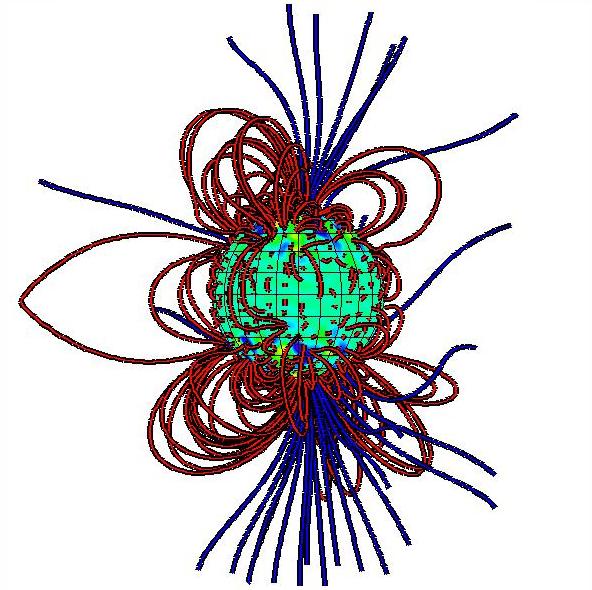}
\hspace{20mm}
\includegraphics[width=55mm]{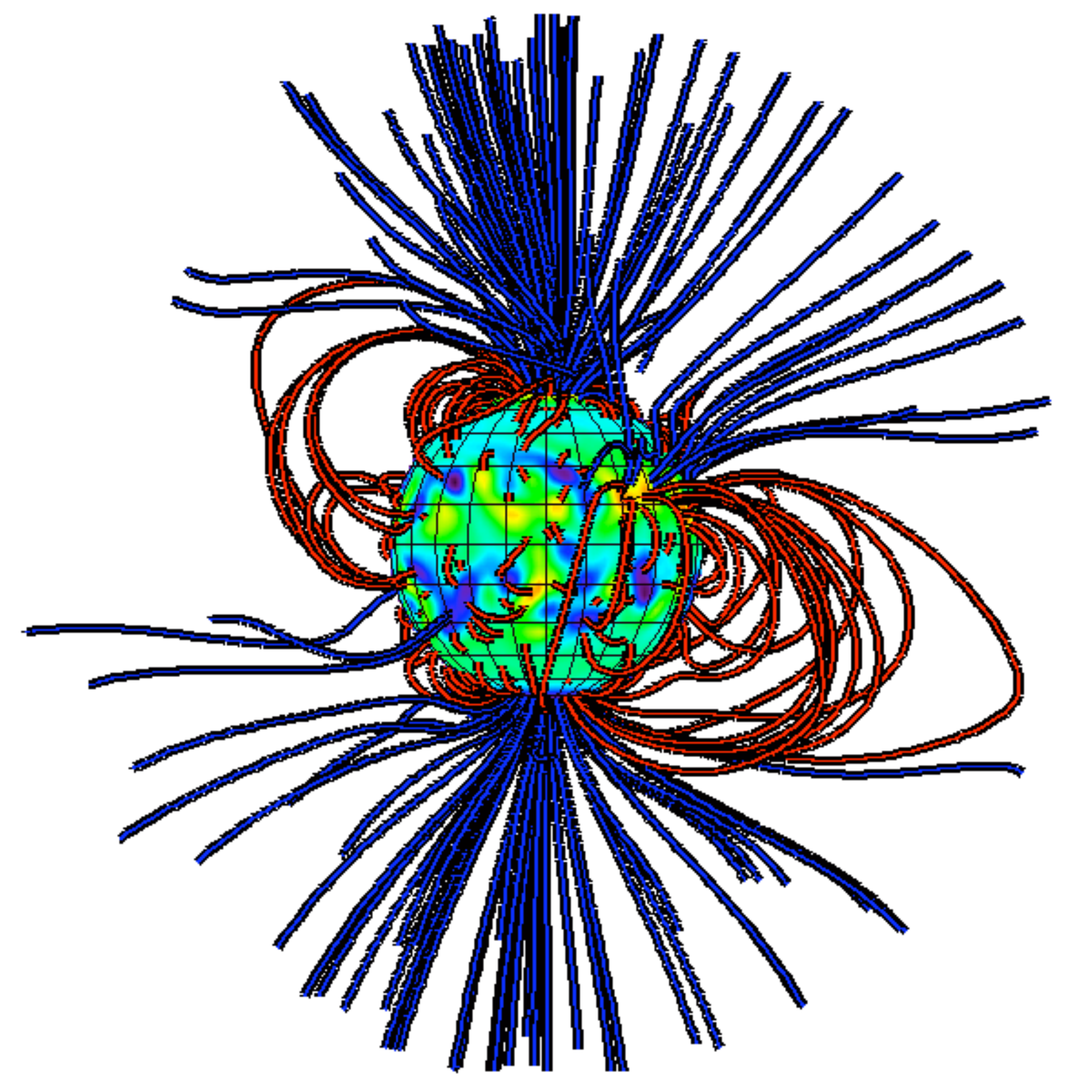}
\includegraphics[width=55mm]{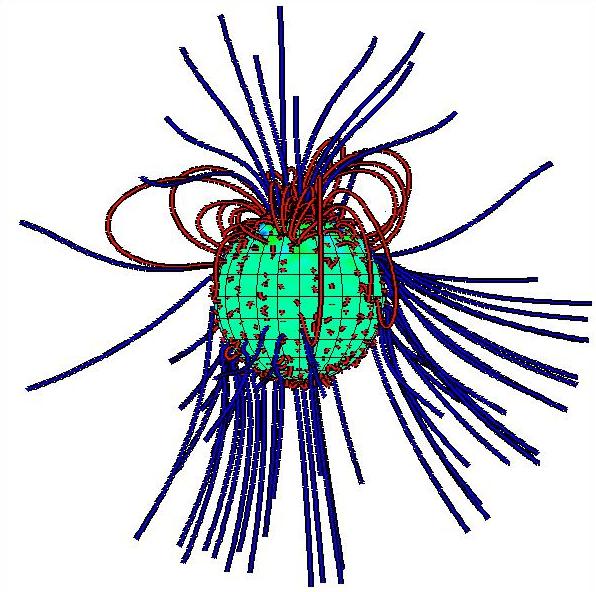}
\hspace{20mm}
\includegraphics[width=55mm]{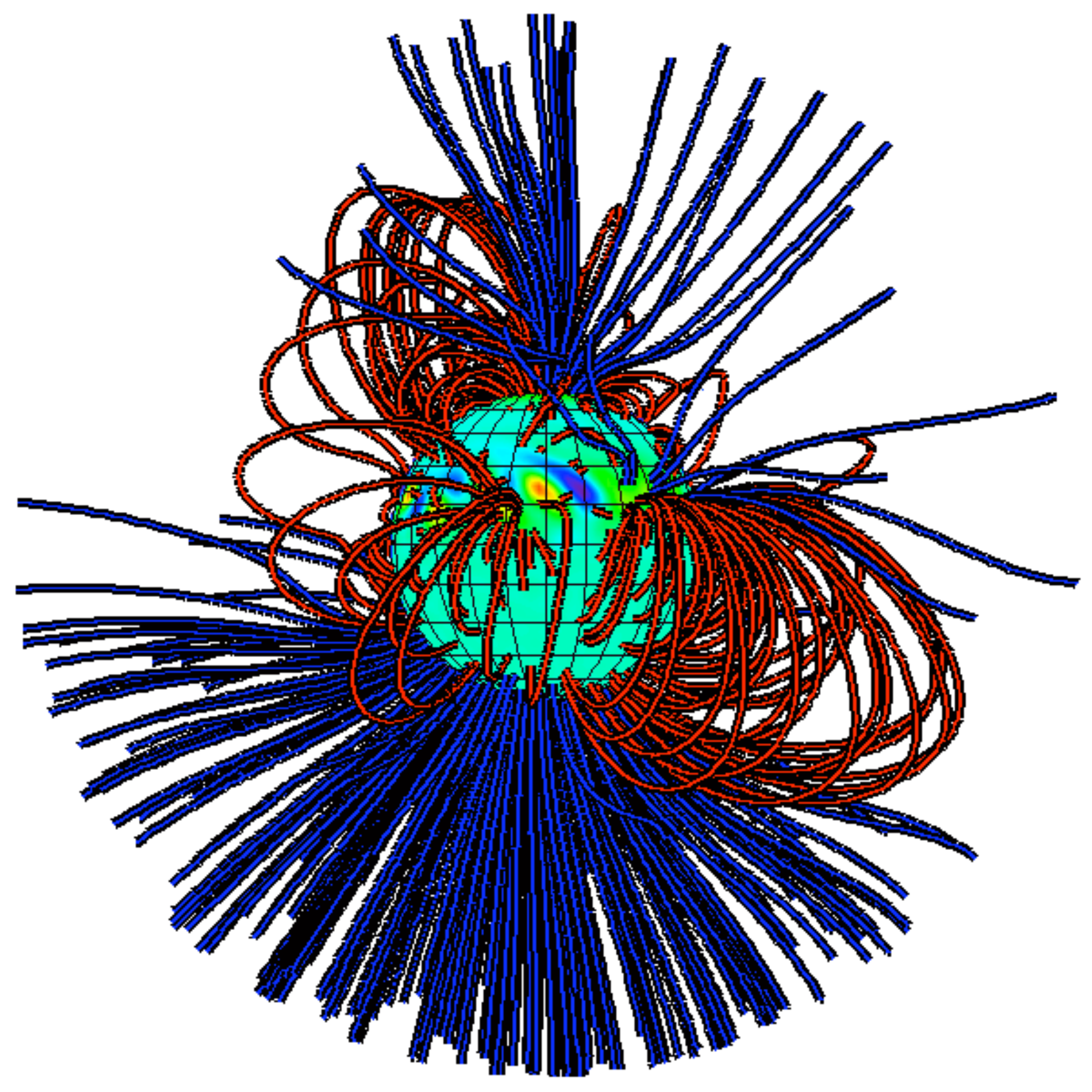}
\includegraphics[width=55mm]{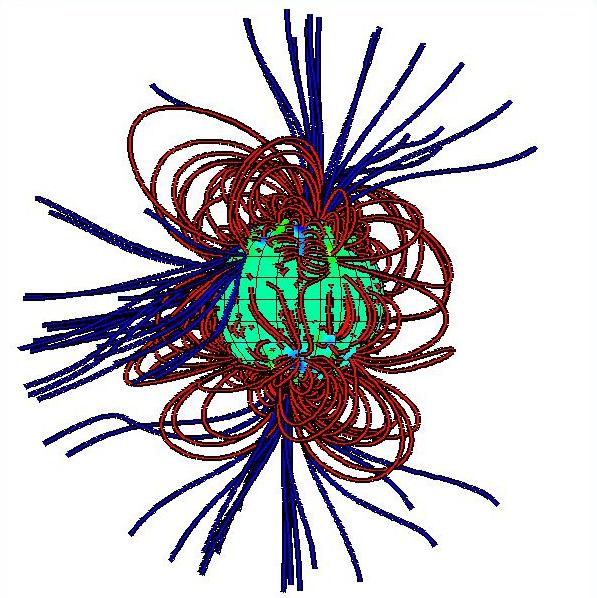}
\hspace{20mm}
\includegraphics[width=55mm]{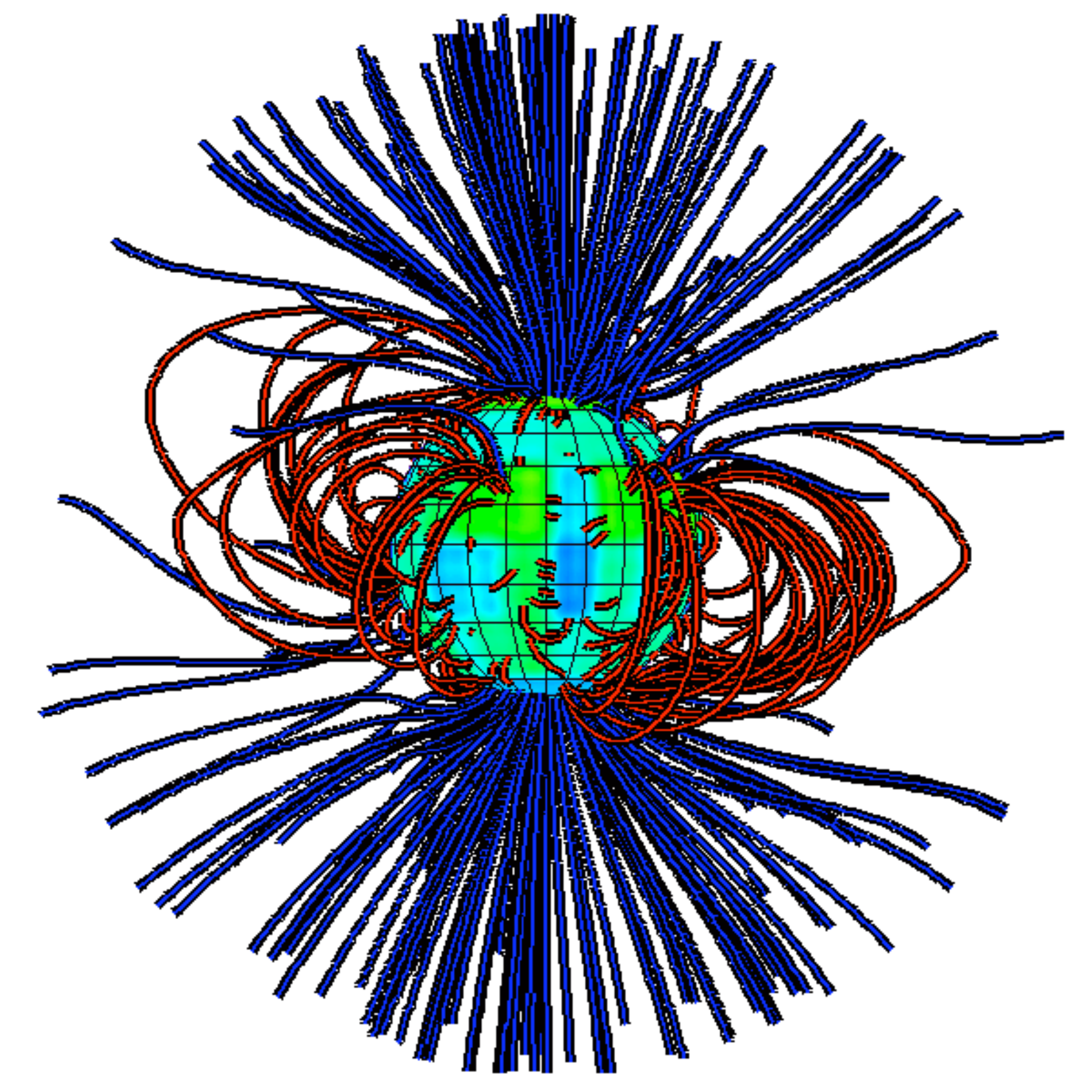}
\caption{Extrapolated coronal fields for the active star (right) and the inactive star (left). Presented for the full data (first row), the field cut off above 500G (second row), the field cutoff in an entire hemisphere (third row) and the field smeared (fourth row). Open field lines are given in blue and closed field lines in red. These images correspond to extrapolations from the surface field maps given in Fig. \ref{surfacemaps} and thus show the star at a period of maximum magnetic activity.  \label{fieldstructure}}
\end{figure*}


\begin{figure*}
\includegraphics[width=170mm]{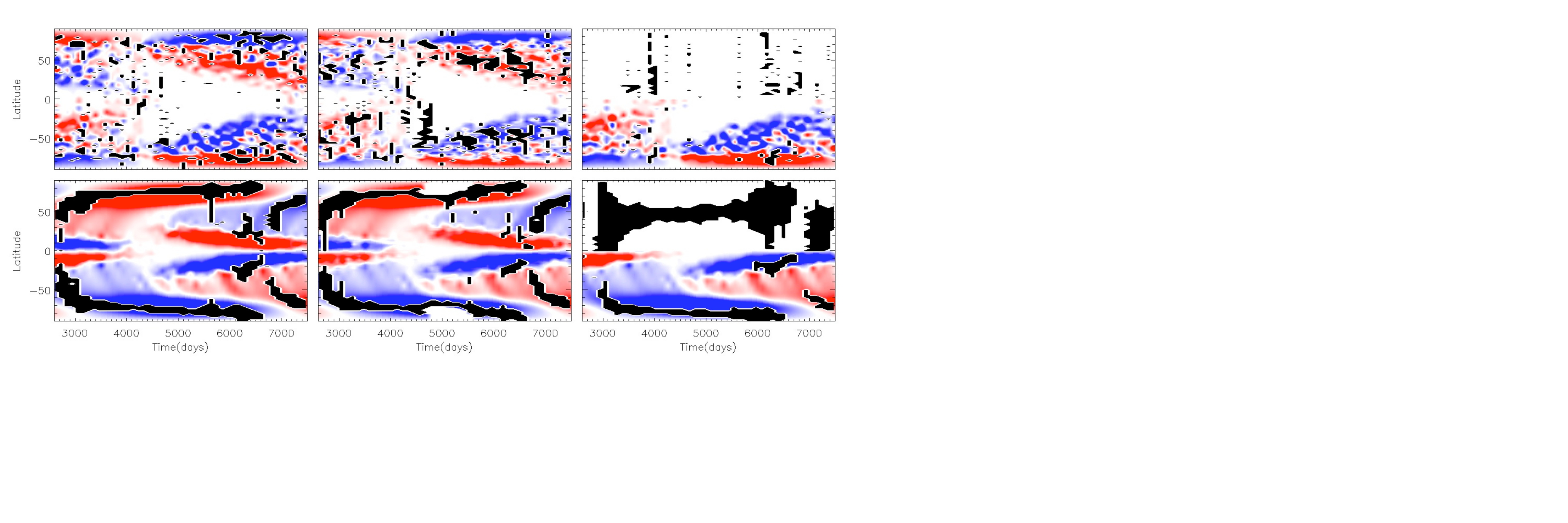}
\caption{Butterfly diagrams for the active star (top row) and inactive star (bottom row) for the full case (left column), suppression of the field stronger than 500G (middle column) and suppression of the field in an entire hemisphere (right column). The black regions show the latitudinal locations where field lines able to support accretion from a circumstellar disc attach to the stellar surface. \label{crossbutterfly}}
\end{figure*}

Extrapolated magnetic fields for both stars are shown in Fig. \ref{fieldstructure}. These diagrams give the field structure at a time of maximum activity corresponding to the surface maps given in Fig. \ref{surfacemaps}.

One of the reasons why understanding the structure of stellar magnetic fields is important is that it will help to understand magnetospheric accretion to young pre-main-sequence stars from their circumstellar discs. In this section, we will discuss the consequences of missing information in this paper has on calculated accretion filling factors and the latitudes at which accreting field lines attach to the stellar surface. For both stars and at all times, the effect of low resolution surface magnetograms is insignificant and will not be discussed here. 

The locations on the stellar surface where accretion can take place are calculated by tracing field lines from every grid point on the stellar surface. For these calculations, the global magnetic fields are extrapolated assuming a source radius of $10 R_*$. Only the field lines that cross the equator and extend to radii between $0.5 R_{co}$ and $1.0 R_{co}$, where $R_{co}$ is the equatorial corotation radius, are assumed to be able to support accretion. All of these field lines are assumed to be supporting accretion. In this case, $R_{co}$ (given by $R_{co}= (G M_* / \omega^2 )^{1/3}$) is calculated assuming the star has solar mass and radius and a rotation period of 2.5 days and has a value of approximately $8 R_*$.

Fig. \ref{crossbutterfly} gives a butterfly diagram showing the latitudes at which field lines that can support accretion connect with the stellar surface over most of an activity cycle. When the field with strengths above 500G are suppressed, it can be seen that there is only a small effect on the accretion filling factors for the inactive star. The active star shows small increases at some times during the cycle with an increase in the amount of low latitude regions contributing to accretion at most times in the cycle. The most significant effect on the accretion filling factors and latitudes comes in the case where the field is suppressed in an entire hemisphere. This is clear for both stars and is due to the fact that without regions of stronger and weaker field in the invisible hemisphere, field lines that are being traced from the visible hemisphere can connect to anywhere with no preferred location. However, when an entire hemisphere is out of view, the accretion pattern in the visible hemisphere tends to remain the same.

Another important aspect that the magnetic structure can have on accretion is its effect on the magnetic pressure in the equatorial plane, which determines the fields ability to disrupt the circumstellar disc. Fortunately, it has been found that for both stars, the missing information causes no significant change in the strength of the magnetic field in the equatorial plane, and its variation as a function of distance from the surface.

Finally, the effect that the missing information has on calculated open magnetic flux will be explored briefly. The two important aspects that will be discussed are the fraction of open flux to total flux, and the latitudinal distribution. These two aspects can affect the rate at which stars lose angular momentum through stellar winds.

In every case, the missing information causes a large drop in the total open flux due to the drop in total unsigned flux. Table \ref{tbl:openflux} gives the average percentage drop in the fraction of open flux to total flux for each case. It can be seen that in almost every case the missing information causes an increase in the fraction of open flux to total flux. The effect of the missing information not only decreases the open flux and the total flux, but also changes the nature of that flux. Typically the flux that is lost is closed. The loss of information leads to a greater fraction of the flux through the stellar surface being open.

For both the active and inactive stars, the open flux tends to be located at the stars' poles. The latitudinal distribution of the open flux is not generally affected by the missing information for both stars. An exception to this is the case of the suppression of the field stronger than 500G on the active star. In all other cases, the open flux is generally located around the poles, whereas in this case, the open flux is generally located at slightly lower latitudes.


\begin{table}
\begin{tabular}{cccc}
\hline
Star &  $\Delta f_{\textrm{spots}}$ (\%) & $\Delta f_{\textrm{hemisphere}}$ (\%) & $\Delta f_{\textrm{smeared}}$ (\%)    \\
\hline
Active & 15 & 26 & 48 \\
Inactive & 9 & -2 & 20 \\
\hline
\end{tabular}
\caption{Percentage increase, averaged over one activity cycle, in the fraction of open flux to total flux, $f=\Delta (\Phi_{open} /\Phi_{total}$), for fields with missing information due to starspots, an entire hemisphere out of view and a limited resolution for both active and inactive stars.} 
\label{tbl:openflux}
\end{table}


\subsection{The coronal density and emission measure}

\begin{table}
\begin{tabular}{cccc}
\hline
Star &  $\Delta n_{e,\textrm{spots}}$ (\%) & $\Delta n_{e,\textrm{hemisphere}}$ (\%)  & $\Delta n_{e,\textrm{smear}}$ (\%)   \\
\hline
Active & -81 & 15 & -35 \\
Inactive & -68 & -1 & -45\\
\hline
\end{tabular}
\caption{Percentage increase, averaged over one activity cycle, in coronal electron density, $n_e$, averaged over the volume of plasma in closed magnetic loops. Results given for both stars due to the absence of magnetic fields in starspots, a hemisphere out of view and the effects of having a surface magnetogram with a reduced resolution.}
\label{tbl:dens}
\end{table}

\begin{figure*}
\includegraphics[width=170mm]{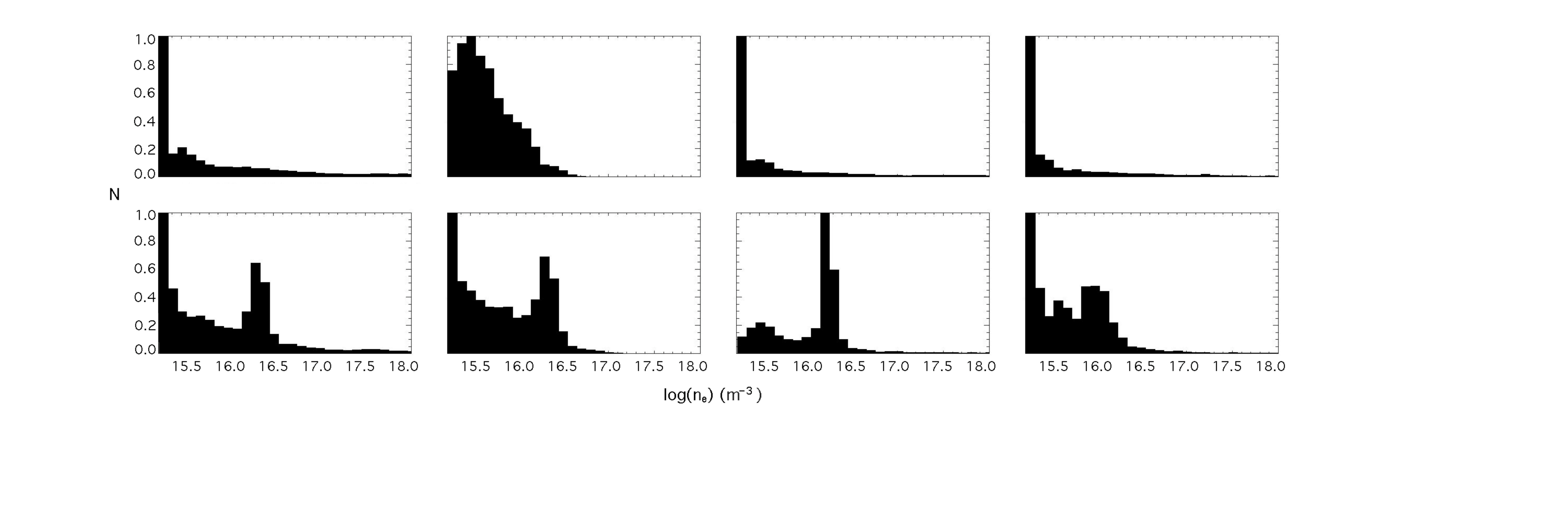}
\caption{Histograms showing the number of regions in the corona with specific electron densities for the active star (top row) and the inactive star (bottom row). The cases of the full data case (first column), the field cutoff above 500G (second column), the field cutoff in an entire hemisphere (third column) and the field smeared (forth column) are given.\label{electrondens}}
\end{figure*}

As described in section \ref{sec:potentialmodel} the density at each point in the corona is determined assuming isothermal, hydrostatic balance along the field line through that point. The one free parameter in this process is the pressure at the base of each field line. Since this is assumed to be proportional to the magnetic pressure there, any degrading of the surface magnetic map will affect the modelled coronal pressure and hence the modeled density. This can be seen clearly in Fig. \ref{electrondens}, which shows histograms of the number of volume elements in the corona at specific electron densities for a time close to cycle maximum (corresponding to Fig. \ref{surfacemaps} and Fig. \ref{structure}). The results are presented for both stars and all cases under consideration. The most significant impact on the distribution of pressures and hence densities in the corona is for the case where the spot fields are removed. Since this cuts out the regions of strongest field, it also cuts out the highest density values, as can be seen by the loss of the high-density tail in the second column of Fig. \ref{electrondens}. 
This would be detected observationally as an enhancement of the densities derived from X-ray lines (e.g. \citealt{2004ApJ...617..508T}) compared to the model values (for the same emission measures). The fraction by which the models would underestimate the densities as a result of the loss of information in the magnetograms is shown in Table \ref{tbl:dens}. It can be seen that in the cases of missing information due to starspots and a limited resolution, large decreases in the average electron densities are present. In the case of missing information in an entire hemisphere, there is no large decrease in the electron densities with the inactive star remaining constant and the active star showing a significant increase.

\begin{figure*}
\includegraphics[width=170mm]{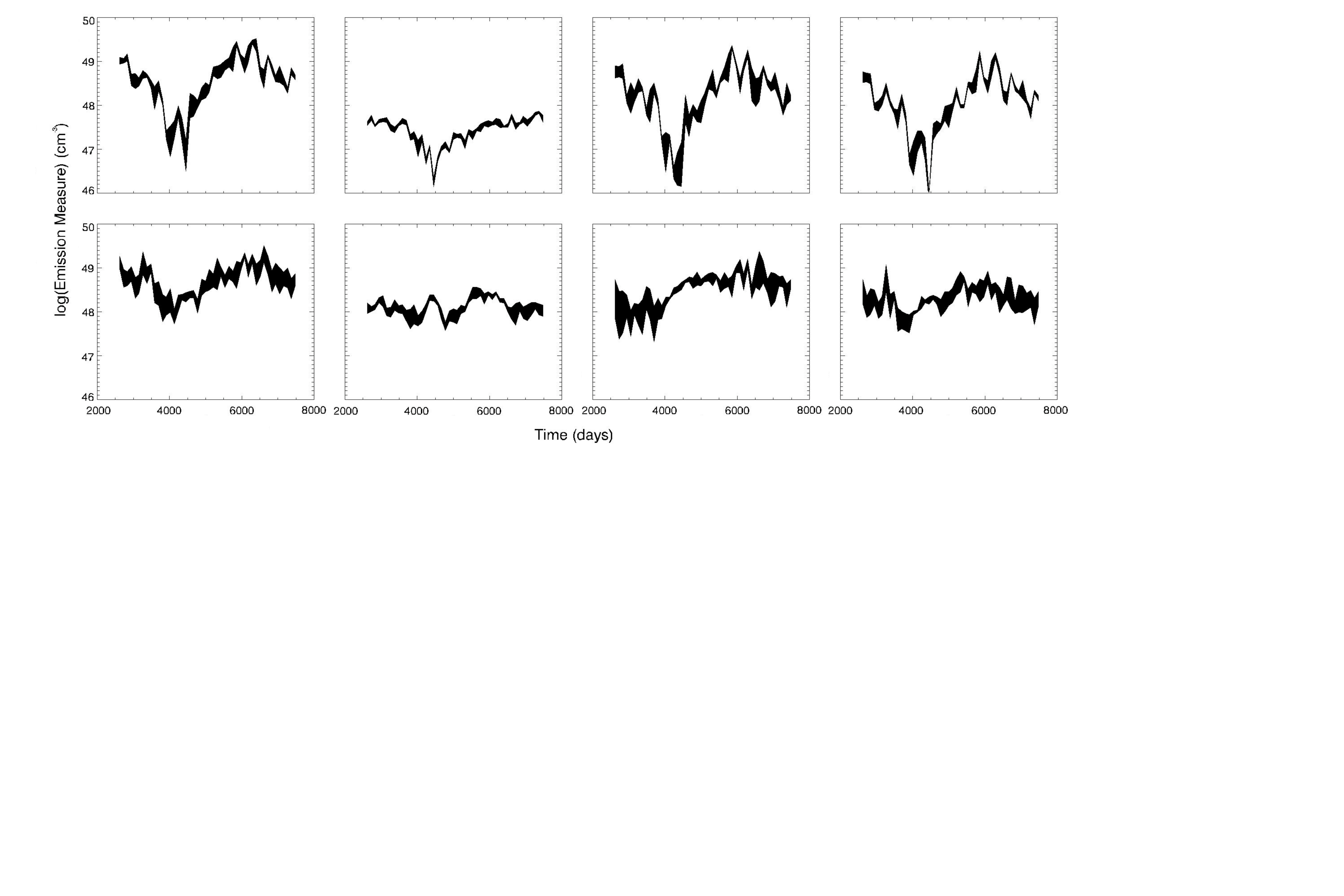}
\caption{Emission measures over one 11-year activity cycle for the active star (top row) and inactive star (bottom row). Results are shown for the full data case (first column), the field cutoff above 500G (second column), the field cutoff in an entire hemisphere (third column) and the field smeared (forth column). The thickness of the lines show  the rotational modulation with the tops of the lines corresponding to the maximum and the bottoms of the lines corresponding to the minimum in visible emission measures as the stars rotate.  \label{em}}
\end{figure*}

\begin{figure}
\includegraphics[width=85mm]{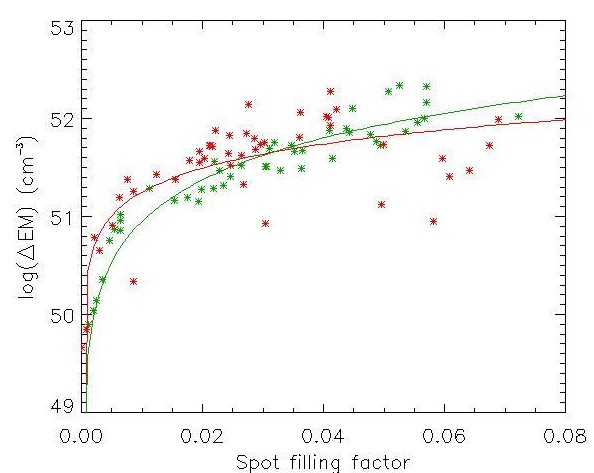}
\caption{Drop in calculated emission measure for both active and inactive stars against the spot filling factor (fraction of stellar surface area with field above 500G) for the active star (green) and inactive star (red). Separate data points correspond to different times over the stellar activity cycle. The thin green line is a best fit to the data for the active star and is given by $\log(\Delta EM) = 53.95 f_s^{0.01}-0.11$. The thin red line is for the inactive star and is given by $\log(\Delta EM) = 53.42 f_s^{0.01} -0.53$. \label{demvsff}}
\end{figure}

This alteration of the modeled coronal density is a response to changes in the surface magnetograms and is also apparent when the emission measures are calculated. Fig. \ref{em} shows the variations of emission measures and the rotational modulation of the emission measures for just over one 11-year activity cycle for both stars. In all cases, the variation in emission measure closely follows the variation in total magnetic flux over the activity cycle shown in Table \ref{tbl:flux}. For the active star, the variation of the emission measure is fairly robust and is only significantly affected by the removal of the field in the spots shown in the second column. The very large drop in the emission measure is due to the large fraction of the surface covered in fields greater than 500G. The inactive star has many fewer high-field regions and consequently a much larger rotational modulation. It is also much more severely affected by both the removal of the spot fields and also by the removal of the field in the lower hemisphere. In both these cases, the form of the variation is changed (indeed the cyclic nature of the variation is all but lost when the spot fields are removed). For both the active and inactive star, the smearing of the maps has least effect, while the removal of the spot fields has the greatest effect. It is only for the inactive star, however, the the form of the cycle is changed. This suggests that calculating stellar coronal emission measures from ZDI maps may fail to reproduce the cyclic variations in emission measures for some stars even if magnetic cycles are present.

The large drops in the modeled emission measures due to the suppression of the strongest fields from the surface magnetograms clearly depends on the spot filling factors. This is shown in Fig. \ref{demvsff} for both stars. It can be seen that the dependance on spot filling factors is the same for both stars.

\section{Conclusion}

A potential field source surface model has been used to extrapolate the global coronal magnetic fields from surface radial field maps of two simulated stars. Using these extrapolated global fields, the coronal emission measures and rotational modulations were calculated over 11-year activity cycles. The calculations were repeated for several cases to simulate the absence of information due to starspots, limited spatial resolution and limited surface visibility due to a tilt in a star's rotation axis. The results were compared in order to understand the limitations of reconstructing coronal magnetic fields in the absence of this information. In all cases, the field structures and emission measures were affected to various degrees. 

In the presence of the simulated starspots, the global flux levels calculated from the surface magnetograms are reduced, but the field structure is not drastically affected on large scales. The largest change in the field comes on a smaller scale from the loss of the most active regions. These results suggest that the presence of starspots does not significantly affect the extrapolation of large scale coronal field but is relevant for studying smaller scale field structures. This does however lead to the largest decrease in the emission measure from the original case of the three effects studied, hinting that starspots may be one of the largest problems when reproducing stellar emission measures. The rotational modulation of emission measures are also significantly affected when the strongest field regions are lost.

The loss of field measurements in dark spotted regions has the greatest impact on the gas pressures inferred from the magnetic field extrapolation. In particular, it reduces the maximum gas pressures (and hence densities) that can be predicted.

Using the doppler imaging technique  \citep{1983ApJ...275..661V}, it is possible to determine the positions and sizes of starspots of sufficient sizes on the surfaces of stars. One possible solution to the problems that starspots introduce may be to artificially impose field onto the surface magnetograms. Another possible solution may be to detect the fields within starspots directly. Using molecular lines that are only formed in starspot umbrae \citep{2002AN....323..192B}, \citet{2008ASPC..384..175B} was able to detect Stokes V signals on at least four M dwarfs. 

The missing information due to a lack of visibility of an entire hemisphere has a greater effect on the final field structure than that of the loss of high field strength regions. In this case, it appeares that the field is weaker and more spread out over different spherical harmonic components. The estimated coronal gas pressures and emission measures are not significantly affected thought in the stellar activity cycle and on average half of the high pressure regions of the stellar corona are preserved. However, the emission measures drop significantly on the rare occasions during the stellar cycle when most of the magnetic flux happens to be in the hemisphere out of view. This is a problem for estimating the emission measure from real stars as one will not be able to know how much magnetic flux covers the missing regions of the observed star.

The observed drop in emission measure for both these cases is due to a loss of the highest coronal gas pressure regions which fill the volume above the high field strength regions. The modeled gas pressures depend on the assumption that the pressure at the base of closed field lines is proportional to the magnetic pressure ($p_0 \propto B_0^2$). Due to this assumption the variations in gas pressures and emission measures for both stars follow closely the total magnetic flux. If a weaker dependance on the magnetic pressure is taken ($p_0 \propto B_0^\alpha$ where $\alpha < 2$), then the dependance of the emission measure on the regions of high field strength will be reduced. Of course a stronger dependance ($\alpha > 2$) would have the opposite effect.

The final case of taking into account the finite resolution limitation on observational reconstructions of surface magnetic fields was shown to have a simplifying effect on the extrapolated field. This suggests that the resolution is only an issue when studying highly complex fields. 

It can be seen in section \ref{sect:fieldstructure} that calculations involving field lines that cross the equator and extend to a radius that could coincide with the inner region of a circumstellar disc can be affected in various ways. When the field in an entire hemisphere is suppressed, the modeled locations of hotspots are largely unaffected in the opposite hemisphere. We are unable to reproduce the location of hotspots in the hemisphere that is not visible leading to an increase in the calculated surface filling factors of hotspots in that hemisphere. In order to study accretion onto stars using field structures extrapolated from ZDI maps, these results suggest that it is better to reproduce plausible field configurations in the hemisphere out of view (as in \citealt{2006MNRAS.371..999G}) although it is possible to reproduce the location and distribution of accretion hotspots in the visible hemisphere, even with incomplete surface magnetograms from ZDI maps.

\section{Acknowledgments}

The authors would like to acknowledge JF Donati and S Gregory for some useful comments. CJ acknowledges support from an STFC studentship.

\bibliographystyle{mn2e}
\bibliography{mybib}

\begin{thebibliography}{}

\bibitem[\protect\citeauthoryear{{Altschuler} \& {Newkirk}}{{Altschuler} \&
  {Newkirk}}{1969}]{1969SoPh....9..131A}
{Altschuler} M.~D.,  {Newkirk} G.,  1969, Solar Phys., 9, 131

\bibitem[\protect\citeauthoryear{{Babcock}}{{Babcock}}{1947}]{1947ApJ...105..1%
05B}
{Babcock} H.~W.,  1947, ApJ, 105, 105

\bibitem[\protect\citeauthoryear{{Berdyugina}}{{Berdyugina}}{2002}]{2002AN....%
323..192B}
{Berdyugina} S.~V.,  2002, Astronomische Nachrichten, 323, 192

\bibitem[\protect\citeauthoryear{{Berdyugina}}{{Berdyugina}}{2005}]{2005LRSP..%
..2....8B}
{Berdyugina} S.~V.,  2005, Living Reviews in Solar Physics, 2, 8

\bibitem[\protect\citeauthoryear{{Berdyugina}, {Fluri}, {Afram} \&
  {Suwald}}{{Berdyugina} et~al.}{2008}]{2008ASPC..384..175B}
{Berdyugina} S.~V.,  {Fluri} D.~M.,  {Afram} N.,    {Suwald} F.,  2008, in
  {G.~van Belle} ed., 14th Cambridge Workshop on Cool Stars, Stellar Systems,
  and the Sun Vol.~384 of Astronomical Society of the Pacific Conference
  Series.
p.~175

\bibitem[\protect\citeauthoryear{{Brown}, {Donati}, {Rees} \& {Semel}}{{Brown}
  et~al.}{1991}]{1991A&A...250..463B}
{Brown} S.~F.,  {Donati} J.-F.,  {Rees} D.~E.,    {Semel} M.,  1991, A\&A, 250,
  463

\bibitem[\protect\citeauthoryear{{Camenzind}}{{Camenzind}}{1990}]{1990RvMA....%
3..234C}
{Camenzind} M.,  1990, in {Klare} G.,  ed., Reviews in Modern Astronomy Vol.~3
  of Reviews in Modern Astronomy.
pp 234--265

\bibitem[\protect\citeauthoryear{{Donati}, {Cameron}, {Semel}, {Hussain},
  {Petit}, {Carter}, {Marsden}, {Mengel}, {L{\'o}pez Ariste}, {Jeffers} \&
  {Rees}}{{Donati} et~al.}{2003}]{2003MNRAS.345.1145D}
{Donati} J.-F.,  {Cameron} A.~C.,  {Semel} M.,  {Hussain} G.~A.~J.,  {Petit}
  P.,  {Carter} B.~D.,  {Marsden} S.~C.,  {Mengel} M.,  {L{\'o}pez Ariste} A.,
  {Jeffers} S.~V.,    {Rees} D.~E.,  2003, MNRAS, 345, 1145

\bibitem[\protect\citeauthoryear{{Donati} \& {Collier Cameron}}{{Donati} \&
  {Collier Cameron}}{1997}]{1997MNRAS.291....1D}
{Donati} J.-F.,  {Collier Cameron} A.,  1997, MNRAS, 291, 1

\bibitem[\protect\citeauthoryear{{Donati}, {Collier Cameron}, {Hussain} \&
  {Semel}}{{Donati} et~al.}{1999}]{1999MNRAS.302..437D}
{Donati} J.-F.,  {Collier Cameron} A.,  {Hussain} G.~A.~J.,    {Semel} M.,
  1999, MNRAS, 302, 437

\bibitem[\protect\citeauthoryear{{Donati}, {Forveille}, {Cameron}, {Barnes},
  {Delfosse}, {Jardine} \& {Valenti}}{{Donati}
  et~al.}{2006}]{2006Sci...311..633D}
{Donati} J.-F.,  {Forveille} T.,  {Cameron} A.~C.,  {Barnes} J.~R.,  {Delfosse}
  X.,  {Jardine} M.~M.,    {Valenti} J.~A.,  2006, Science, 311, 633

\bibitem[\protect\citeauthoryear{{Donati}, {Howarth}, {Jardine}, {Petit},
  {Catala}, {Landstreet}, {Bouret}, {Alecian}, {Barnes}, {Forveille}, {Paletou}
  \& {Manset}}{{Donati} et~al.}{2006}]{2006MNRAS.370..629D}
{Donati} J.-F.,  {Howarth} I.~D.,  {Jardine} M.~M.,  {Petit} P.,  {Catala} C.,
  {Landstreet} J.~D.,  {Bouret} J.-C.,  {Alecian} E.,  {Barnes} J.~R.,
  {Forveille} T.,  {Paletou} F.,    {Manset} N.,  2006, MNRAS, 370, 629

\bibitem[\protect\citeauthoryear{{Donati}, {Jardine}, {Gregory}, {Petit},
  {Bouvier}, {Dougados}, {M{\'e}nard}, {Cameron}, {Harries}, {Jeffers} \&
  {Paletou}}{{Donati} et~al.}{2007}]{2007MNRAS.380.1297D}
{Donati} J.-F.,  {Jardine} M.~M.,  {Gregory} S.~G.,  {Petit} P.,  {Bouvier} J.,
   {Dougados} C.,  {M{\'e}nard} F.,  {Cameron} A.~C.,  {Harries} T.~J.,
  {Jeffers} S.~V.,    {Paletou} F.,  2007, MNRAS, 380, 1297

\bibitem[\protect\citeauthoryear{{Donati}, {Jardine}, {Gregory}, {Petit},
  {Paletou}, {Bouvier}, {Dougados}, {M{\'e}nard}, {Cameron}, {Harries},
  {Hussain}, {Unruh}, {Morin}, {Marsden}, {Manset}, {Auri{\`e}re}, {Catala} \&
  {Alecian}}{{Donati} et~al.}{2008}]{2008MNRAS.386.1234D}
{Donati} J.-F.,  {Jardine} M.~M.,  {Gregory} S.~G.,  {Petit} P.,  {Paletou} F.,
   {Bouvier} J.,  {Dougados} C.,  {M{\'e}nard} F.,  {Cameron} A.~C.,  {Harries}
  T.~J.,  {Hussain} G.~A.~J.,  {Unruh} Y.,  {Morin} J.,  {Marsden} S.~C.,
  {Manset} N.,  {Auri{\`e}re} M.,  {Catala} C.,    {Alecian} E.,  2008, MNRAS,
  386, 1234

\bibitem[\protect\citeauthoryear{{Donati}, {Wade}, {Babel}, {Henrichs}, {de
  Jong} \& {Harries}}{{Donati} et~al.}{2001}]{2001MNRAS.326.1265D}
{Donati} J.-F.,  {Wade} G.~A.,  {Babel} J.,  {Henrichs} H.~f.,  {de Jong}
  J.~A.,    {Harries} T.~J.,  2001, MNRAS, 326, 1265

\bibitem[\protect\citeauthoryear{{Gregory}, {Jardine}, {Simpson} \&
  {Donati}}{{Gregory} et~al.}{2006}]{2006MNRAS.371..999G}
{Gregory} S.~G.,  {Jardine} M.,  {Simpson} I.,    {Donati} J.-F.,  2006, MNRAS,
  371, 999

\bibitem[\protect\citeauthoryear{{Hussain}, {Collier Cameron}, {Jardine},
  {Dunstone}, {Ramirez Velez}, {Stempels}, {Donati}, {Semel}, {Aulanier},
  {Harries}, {Bouvier}, {Dougados}, {Ferreira}, {Carter} \& {Lawson}}{{Hussain}
  et~al.}{2009}]{2009arXiv0905.0914H}
{Hussain} G.~A.~J.,  {Collier Cameron} A.,  {Jardine} M.~M.,  {Dunstone} N.,
  {Ramirez Velez} J.,  {Stempels} H.~C.,  {Donati} J.~.,  {Semel} M.,
  {Aulanier} G.,  {Harries} T.,  {Bouvier} J.,  {Dougados} C.,  {Ferreira} J.,
  {Carter} B.~D.,    {Lawson} W.~A.,  2009, ArXiv e-prints

\bibitem[\protect\citeauthoryear{{Hussain}, {Jardine}, {Collier Cameron},
  {Barnes} \& {Donati}}{{Hussain} et~al.}{2000}]{2000ASPC..198..463H}
{Hussain} G.~A.~J.,  {Jardine} M.~M.,  {Collier Cameron} A.,  {Barnes} J.~R.,
   {Donati} J.-F.,  2000, in {Pallavicini} R.,  {Micela} G.,   {Sciortino} S.,
  eds, Stellar Clusters and Associations: Convection, Rotation, and Dynamos
  Vol.~198 of Astronomical Society of the Pacific Conference Series.
p.~463

\bibitem[\protect\citeauthoryear{{Jardine}, {Barnes}, {Donati} \& {Collier
  Cameron}}{{Jardine} et~al.}{1999}]{1999MNRAS.305L..35J}
{Jardine} M.,  {Barnes} J.~R.,  {Donati} J.-F.,    {Collier Cameron} A.,  1999,
  MNRAS, 305, L35

\bibitem[\protect\citeauthoryear{{Jardine}, {Collier Cameron} \&
  {Donati}}{{Jardine} et~al.}{2002}]{2002MNRAS.333..339J}
{Jardine} M.,  {Collier Cameron} A.,    {Donati} J.-F.,  2002, MNRAS, 333, 339

\bibitem[\protect\citeauthoryear{{K$\ddot{o}$enigl}}{{K$\ddot{o}$enigl}}{1991}%
]{1991ApJ...370L..39K}
{K$\ddot{o}$enigl} A.,  1991, ApJ, 370, L39

\bibitem[\protect\citeauthoryear{{Long}, {Romanova} \& {Lovelace}}{{Long}
  et~al.}{2008}]{2008MNRAS.386.1274L}
{Long} M.,  {Romanova} M.~M.,    {Lovelace} R.~V.~E.,  2008, MNRAS, 386, 1274

\bibitem[\protect\citeauthoryear{{Mackay}, {Jardine}, {Cameron}, {Donati} \&
  {Hussain}}{{Mackay} et~al.}{2004}]{2004MNRAS.354..737M}
{Mackay} D.~H.,  {Jardine} M.,  {Cameron} A.~C.,  {Donati} J.-F.,    {Hussain}
  G.~A.~J.,  2004, MNRAS, 354, 737

\bibitem[\protect\citeauthoryear{{Morin}, {Donati}, {Forveille}, {Delfosse},
  {Dobler}, {Petit}, {Jardine}, {Cameron}, {Albert}, {Manset}, {Dintrans},
  {Chabrier} \& {Valenti}}{{Morin} et~al.}{2008}]{2008MNRAS.384...77M}
{Morin} J.,  {Donati} J.-F.,  {Forveille} T.,  {Delfosse} X.,  {Dobler} W.,
  {Petit} P.,  {Jardine} M.~M.,  {Cameron} A.~C.,  {Albert} L.,  {Manset} N.,
  {Dintrans} B.,  {Chabrier} G.,    {Valenti} J.~A.,  2008, MNRAS, 384, 77

\bibitem[\protect\citeauthoryear{{Morin}, {Donati}, {Petit}, {Delfosse},
  {Forveille}, {Albert}, {Auri{\`e}re}, {Cabanac}, {Dintrans}, {Fares},
  {Gastine}, {Jardine}, {Ligni{\`e}res}, {Paletou}, {Ramirez Velez} \&
  {Th{\'e}ado}}{{Morin} et~al.}{2008}]{2008MNRAS.390..567M}
{Morin} J.,  {Donati} J.-F.,  {Petit} P.,  {Delfosse} X.,  {Forveille} T.,
  {Albert} L.,  {Auri{\`e}re} M.,  {Cabanac} R.,  {Dintrans} B.,  {Fares} R.,
  {Gastine} T.,  {Jardine} M.~M.,  {Ligni{\`e}res} F.,  {Paletou} F.,  {Ramirez
  Velez} J.~C.,    {Th{\'e}ado} S.,  2008, MNRAS, 390, 567

\bibitem[\protect\citeauthoryear{{Romanova}, {Koldoba}, {Ustyugova},
  {Kulkarni}, {Long} \& {Lovelace}}{{Romanova}
  et~al.}{2009}]{2009ASPC..406..112R}
{Romanova} M.~M.,  {Koldoba} A.~V.,  {Ustyugova} G.~V.,  {Kulkarni} A.~K.,
  {Long} M.,    {Lovelace} R.~V.~E.,  2009, in {Pogorelov} N.~V.,  {Audit} E.,
  {Colella} P.,   {Zank} G.~P.,  eds, Astronomical Society of the Pacific
  Conference Series Vol.~406.
p.~112

\bibitem[\protect\citeauthoryear{{Semel}}{{Semel}}{1989}]{1989A&A...225..456S}
{Semel} M.,  1989, A\&A, 225, 456

\bibitem[\protect\citeauthoryear{{Solanki}}{{Solanki}}{2002}]{2002AN....323..1%
65S}
{Solanki} S.~K.,  2002, Astronomische Nachrichten, 323, 165

\bibitem[\protect\citeauthoryear{{Testa}, {Drake} \& {Peres}}{{Testa}
  et~al.}{2004}]{2004ApJ...617..508T}
{Testa} P.,  {Drake} J.~J.,    {Peres} G.,  2004, ApJ, 617, 508

\bibitem[\protect\citeauthoryear{{van Ballegooijen}, {Cartledge} \&
  {Priest}}{{van Ballegooijen} et~al.}{1998}]{1998ApJ...501..866V}
{van Ballegooijen} A.~A.,  {Cartledge} N.~P.,    {Priest} E.~R.,  1998, ApJ,
  501, 866

\bibitem[\protect\citeauthoryear{{Vogt} \& {Penrod}}{{Vogt} \&
  {Penrod}}{1983}]{1983ApJ...275..661V}
{Vogt} S.~S.,  {Penrod} G.~D.,  1983, ApJ, 275, 661

\end{thebibliography}

\end{document}